\documentclass[%
 reprint,
 amsmath,amssymb,
 aps,
]{revtex4-1}

\usepackage{graphicx}
\usepackage{dcolumn}
\usepackage{bm}

\def\be{\begin{equation}}
\def\ee{\end{equation}}
\def\ba{\begin{eqnarray}}
\def\ea{\end{eqnarray}}

\begin{document}

\preprint{APS/123-QED}

\title{CMB Constraints on a Stochastic Background of Primordial Magnetic 
Fields} 

\author{Daniela Paoletti}\email{paoletti@iasfbo.inaf.it}
\affiliation{Dipartimento di Fisica and INFN,universit\`a degli Studi 
di Ferrara,\\
via Saragat, 2 I-44100 Ferrara - Italy}
\affiliation{INAF/IASF-BO,Istituto di Astrofisica Spaziale e Fisica 
Cosmica di Bologna \\
via Gobetti 101, I-40129 Bologna - Italy}

\author{Fabio Finelli}\email{finelli@iasfbo.inaf.it}
\affiliation{INAF/IASF-BO,
Istituto di Astrofisica Spaziale e Fisica
Cosmica di Bologna \\
via Gobetti 101, I-40129 Bologna - Italy}
\affiliation{INFN, Sezione di Bologna,
Via Irnerio 46, I-40126 Bologna, Italy}

\date{\today}
\begin{abstract}
We constrain a stochastic background of primordial magnetic field (PMF) by its contribution to 
the angular power spectrum of 
cosmic microwave background anisotropies. 
We parametrize such stochastic background by a power-law spectrum with index $n_B$ and by its Gaussian smoothed amplitude 
$B_\lambda$ on a comoving length $\lambda$. We give an approximation for the spectra 
of the relevant correlators of the energy-momentum of the stochastic background of PMF for any $n_B$.
By using the WMAP 7 year data in combination with ACBAR, BICEP and QUAD 
we obtain the constraint $B_{1 {\rm Mpc}} < 5.0$ nG at $95\%$ confidence level
for a stochastic background of non-helical PMF.
We discuss the relative importance of the scalar and vector 
contribution to CMB anisotropies in obtaining these constraints. We then forecast {\sc Planck} capabilities in constraining 
$B_{1 {\rm Mpc}}$. 
\end{abstract}

\pacs{Valid PACS appear here}
\maketitle

\section{Introduction}

The origin of the large scale magnetic fields observed in galaxies and clusters of galaxies is an open issue of
great importance in modern astrophysics (see \cite{Widrow:2002ud} for a review).
Primordial magnetic fields (PMF) generated in the early Universe
could have been the seeds for large scale magnetic fields 
and have left an imprint in the
anisotropy pattern of the cosmic microwave background (CMB). 
A primordial hypothesis for generating the seeds amplified afterwards by adiabatic compression and dynamo - 
cannot be discarded \cite{Widrow:2002ud}, also in light of recent observations of strong magnetic fields in 
galaxies at high redshift \cite{Bernet:2008qp,Wolfe:2008nk}. 

PMF with a comoving amplitude of several nG can leave interesting imprints on CMB anisotropies.
A stochastic background of PMF is modelled as a fully inhomogeneous component
and its energy momentum tensor (EMT) - quadratic in the magnetic fields -
is considered at the same footing as linear inhomogeneities in the
other components and linear metric fluctuations.
A stochastic background of PMF generates independent modes for
all kinds of linear perturbations: there has been several studies
for scalar \cite{KL,yamazaki,KR,GK,FPP,PFP,SL,BoCa},
vector \cite{SB,MKK,lewis} and tensor \cite{DFK,MKK,CDK} perturbations. 
See Refs. \cite{PFP,SL} for studies which take into account all the types of contributions. 
A stochastic background of PMF affects also the statistics of CMB anisotropies, and not only its power spectra: 
being quadratic in the magnetic field 
amplitude, the EMT is non-Gaussian distributed \cite{BC} and therefore 
the bispectrum of CMB anisotropies can also be a useful probe \cite{CFPR,SS}.

In our previous works \cite{FPP,PFP} we have refined the computation of CMB anisotropies 
in presence of a stochastic background of PMF: we have 
computed the initial conditions for cosmological perturbations in the radiation era 
keeping into account only relativistic degrees of freedom 
and the correlators for the Fourier transforms of the EMT in presence of a sharp cut-off which mimics  
the damping scale due to viscosity \cite{FPP,PFP} for few values of the spectral index $n_B$ of the stochastic background. 

In this paper we use and extend the above theoretical description 
to derive the CMB constraints on a stochastic background of PMF which can be obtained 
by current and future data.
We therefore use a modified version of \texttt{CosmoMC} \cite{cosmomc} 
connected with a modified version of \texttt{CAMB} \cite{camb} containing all the above features to 
constrain $B_\lambda$ - the amplitude of the magnetic field smoothed over a comoving scale $\lambda$ -
and $n_B$ with the most recent compilation of CMB anisotropies data, 
therefore updating previous investigations \cite{yamazaki,yamazakilast,shawlewis2}. 
We also forecast the {\sc Planck} \cite{bluebook} capabilities in constraining such a background of PMF.

Our paper is organized as follows.
In Section II we present the theoretical set-up needed for deriving the CMB constraints.
As new results we give an approximation for the PMF energy-momentum valid for any 
$n_B$, we discuss the correlator between the energy density and the Lorentz force
and we extend the regular initial conditions for cosmological perturbations in the relativistic regime 
in presence of a stochastic background of PMF \cite{PFP} including the correction due to matter, collecting the details in Appendix A, 
B and C, respectively.  
In Section III we discuss the constraints from WMAP 7 years data \cite{jarosik,larson}, ACBAR \cite{ACBAR}, 
BICEP \cite{BICEP} and QUAD \cite{QUAD} on a flat $\Lambda$CDM model plus a stochastic background of PMF. 
We present the Planck capabilities in constraining $B_{\lambda}$ and $n_B$ in Section IV and 
we summarize our results in Section V.

\section{Stochastic Background of PMF}

A stochastic background of PMF acts as a fully inhomogeneous source to the Einstein equations.
As usual, we assume the infinite conductivity limit in which the
PMF simply scales with time as $B({\bf x},\tau )=B({\bf x})/a(\tau)^2$ where $a(\tau)$ is the scale
factor normalized to $a_0=1$ today and $\tau$ is the conformal time.
We model PMF with a power-law spectrum $P_B(k)=A\, k^{n_B}$.
The two point correlation function for inhomogeneous fields is:
\be
\langle B_i({\bf k}) B_j^*({\bf k}')\rangle=(2\pi)^3 \delta({\bf k}-{\bf k}')
(\delta_{ij}-\hat k_i\hat k_j) \frac{P_B(k)}{2}
\ee 
where $n_B > -3$ to avoid infrared singularities.
Magnetic perturbations survive the Silk damping but on smaller scales PMF are damped by radiation viscosity. 
We model this damping introducing a sharp cut off in the spectrum at a damping scale $k_D$ \cite{SB} which is much smaller 
than the Silk scale.
To parametrize PMF amplitude, we choose to use the most common convention which smooths 
the magnetic fields with a Gaussian filter over a comoving scale $\lambda$: 
\be
B^2_\lambda = \int_0^{\infty} \frac{d{k \, k^2}}{2 \pi^2} e^{-k^2 \lambda^2}
P_B (k) = \frac{A}{4 \pi^2 \lambda^{n_B+3}}
\Gamma \left( \frac{n_B+3}{2} \right)
\label{gaussian}
\ee
As damping scale we adopt the proposal of Ref. \cite{SB}, in which $k_D$ is function
of $B_\lambda$ and the spectral index $n_B$:
\be
k_D=(2.9 \times 10^4)^\frac{1}{n_B+5} \Big(\frac{B_\lambda}{\rm nG}\Big)^\frac{-2}{n_B+5} 
\Big(\frac{2\pi}{\lambda/{\rm Mpc}}\Big)^\frac{n_B+3}{n_B+5}h^\frac{1}{n_B+5}\,.
\label{kd_def}
\ee
The EMT of PMF is a source term for the Einstein-Boltzmann system and, 
being these quadratic in the $B_i({\bf k})$, the Fourier
transfoms of the EMT components are convolutions \cite{MKK,DFK}:
\ba
|\rho_B(k)|^2&=&\frac{1}{1024\pi^5}\int_\Omega d{\bf p} P_B(p) \, P_B(|{\bf k}-{\bf p}|)(1+\mu^2)
\label{density}
\\
|\Pi^{(V)}(k)|^2&=&\frac{1}{512\pi^5}\int_\Omega d{\bf p} P_B(p)\, P_B(|{\bf k}-{\bf p}|)\nonumber\\
&\times&[(1+\beta^2)(1-\gamma^2) + \gamma\beta(\mu-\gamma\beta)]
\label{vector}
\\
|\Pi^{(T)}(k)|^2&=&\frac{1}{512\pi^5} \int_\Omega d{\bf p} P_B(p)P_B(|{\bf k}-{\bf p}|)\nonumber\\
 &\times&(1+2\gamma^2+\gamma^2\beta^2) \,,
\label{tensor}
\ea
where $\mu = \hat {\bf p} \cdot ({\bf k} -{\bf p})/|{\bf k} -{\bf p}|$,
$\gamma= \hat {\bf k} \cdot \hat {\bf p}$,
$\beta= \hat {\bf k} \cdot ({\bf k} -{\bf p})/|{\bf k} -{\bf p}|$ and $\Omega$ denotes the volume with $p<k_D$.\\

The analytical exact results for Eqs. (4-6) are given for specific values of $n_B$ in our previous papers 
\cite{FPP,PFP}. However, the expressions for generic $n_B$ are rather complicated
and cannot be used in a numerical implementation. Therefore we fitted the analytical results
with easier expressions to be inserted as spectra for the PMF in 
modified version of \texttt{CAMB} \cite{camb} and \texttt{CosmoMC} \cite{cosmomc}.
Since the spectral shape varies with the spectral index we had to divide the spectra in different index ranges.
The first natural split is between indices greater and smaller than $n_B=-3/2$:
this division is very natural since is required by the change in the infrared behaviour between the two ranges.
In order to achieve the best accuracy with the fits and considered the wide range 
of spectral indices that we wanted to explore we decided to do a further splitting in 
the $n_B>-3/2$ range between positive and negative spectral indices.
In the end we result into three different spectral fits for each EMT component.
In Fig.\ref{Comparison_fit_scal} we show respectively for scalar energy density, 
scalar Lorentz force and vector anisotropic stress 
the results of our fits compared with the analytical results.
We note how the fits are in excellent agreement with the analytical results and we refer the reader to Appendix A for the details.
In the next two subsections we address other two issues related to the theoretical characterization of a stochastic background of PMF, recently 
addressed in Refs. \cite{SL,Kojima}.
  
\subsection{Cross-Correlation between Lorentz Force and Energy Density}

The first issue is represented by the cross correlation between Lorentz force and magnetic energy density
as pointed out in \cite{SL}.
In particular the analysis of the magnetized scalar mode involves three
quantities: magnetic energy density, Lorentz force and anisotropic stress.
These quantities are not independent one from the other but are related throught
the conservation equation for the PMF EMT (in the magnetic hydrodynamic limit):
$\sigma_B=\frac{\rho_B}{3}+L_B$.
We approximated the Lorentz force and the magnetic energy density as 
anti-correlated \cite{FPP,PFP}, however, the Lorentz force and the magnetic energy density 
correlation can be calculated, as for the their auto-correlations, as pointed 
out in Ref. \cite{SL}.   
The cross-correlator between $\rho_B$ and $L_B$ is given by:
\ba
\langle\rho_B({\bf{k}})L_B({\bf{k'}})\rangle=
\frac{\delta({\bf{k}}-{\bf{k'}})}{1024\pi^5} \int d{\bf p} P_B(p) \, P_B(|{\bf k}-{\bf p}|)\nonumber\\
\times (1-1(\gamma^2+\beta^2)+2 \gamma\beta\mu-\mu^2)
\label{cross}
\ea
The above formula can be computed analitycally and in the appendix we show the analytical results
both for $n_B=-2.5$ and $n_B=2$. The general behavior in the large scale limit of the spectrum 
(for $k<<k_D$), the range relevant for CMB anisotropies, is:
\ba
\langle\rho_B(k)L_B(k)\rangle&=&-\frac{1}{3}|\rho_B(k)|^2 \,\,{\rm for} \, n_B\ge -3/2\nonumber\\
\langle\rho_B(k)L_B(k)\rangle&=& -C|\rho_B(k)|^2 \,\,{\rm for}\, n_B< -3/2\nonumber
\ea
where $C\sim \mathcal{O}(1)$.
In the appendix we show that the exact evaluation of the cross 
correlation has a small effect - in particular for strongly negative 
$n_B$ - with respect to our choice of considering the Lorentz 
force and the energy density as fully anti-correlated. 

\subsection{Magnetized Scalar Initial Conditions with matter Corrections}

The second issue is related with the initial conditions for magnetized scalar perturbations.
In our previous works \cite{FPP,PFP} we computed the intial conditions in the approximation of 
a universe dominated by radiation, following the results presented in Ref. \cite{Kojima}
we investigated the importance of matter contributions to initial conditions.
We computed the initial conditions with the inclusion of matter corrections and we give 
the details in appendix C. The results show that 
the introduction of matter corrections in the initial conditions have a negligible 
impact on the numerical computation of the magnetized scalar CMB anisotropies.

\section{Constraints from current CMB data}


We derived the constraints on PMF with current CMB anisotropy data performing an analysis of the 
WMAP 7 year \cite{jarosik,larson}, ACBAR~\cite{ACBAR} data in temperature
and of the BICEP \cite{BICEP} and QUaD \cite{QUAD} data in polarization. 
In order to decrease the correlations between different data sets which cover the same region of
the sky, we proceed as in Ref. \cite{fhll}.
We remove in the analysis the following CMB band powers: a) all the QUaD TT band powers since they
overlap with data from the `CMB8' region of ACBAR, b) the ACBAR band powers with $\ell<910$ and $\ell>1950$ to
avoid overlap with WMAP (which is cosmic variance limited up to $\ell=919$~\cite{jarosik,larson}) and contamination from
foreground residuals,respectively, c) the QUAD TE band powers which overlap with WMAP ones, the 
QUAD EE band powers which overlap with BICEP, d) the BICEP TT, TE band powers (i.e., we use just EE and BB
information from BICEP).

We use a modified version of \texttt{CosmoMC} \cite{cosmomc} in order to compute 
the Bayesian probability distribution of cosmological parameters, including the magnetic ones.
We vary the baryon density $\omega_{b}=\Omega_{b} h^2$, the cold dark matter density 
$\omega_{c}= \Omega_{c}h^2$ (with $h$ being
$H_0/100 {\rm km}\,{\rm s}^{-1}{\rm Mpc}^{-1}$), the reionisation optical depth 
$\tau$ (not to be confused with the 
conformal time $\tau$), 
the ratio of the sound horizon to the angular diameter distance at decoupling $\theta$, $\ln ( 10^{10} A_S )$, $n_S$ 
and the magnetic parameters $B_{| 1 {\rm Mpc}}$ (in units of $10 {\rm nG}$) and $n_B$. 
As priors we use $\left[0 \,, 10 \right]$ for $B_{1 {\rm Mpc}}/(10 {\rm nG})$ and $\left[ -2.9 \,, 3 \right]$ for $n_B$. 
The damping scale $k_D$ defined in Eq. (\ref{kd_def}) is obtained as a derived parameter (in units of Mpc$^{-1}$), 
as well as $H_0$. 


We assume a flat universe, a CMB temperature $T_{\rm CMB}=2.725$~K and we set the primordial 
Helium fraction to $y_{\rm He}=0.24$. We use the recombination routine RECFAST v1.5 \cite{SSS}.  
We restrict our analysis to three massless neutrinos (a non-vanishing 
neutrino mass leads to a large scale enhancement in the power spectrum of CMB anisotropies in the presence of PMF 
\cite{SL} and would not change our results).
The pivot scale of the primordial scalar was set to
$k_*=0.05$~Mpc$^{-1}$. In order to fit WMAP 7 years, ACBAR and QUaD data, we use the CMB 
power spectra and we follow the method implemented in
\texttt{CosmoMC} consisting in varying a nuisance parameter $A_{\rm
SZ}$ which accounts for the unknown amplitude of the thermal Sunyaev-Zeldovich (SZ)
contribution to the small-scale CMB data points assuming the model of
\cite{KoSe}. We use a \texttt{CAMB} accuracy setting of $1$.
We sample the posterior using the
Metropolis-Hastings algorithm \cite{Hastings:1970xy} at a temperature
$T=2$ (the temperature parameter in the \texttt{CosmoMC} code 
is used to sample the probability $P$ as $P^{1/T}$, allowing a better 
exploration of the distribution tails), generating four 
parallel chains and imposing a conservative Gelman-Rubin convergence
criterion \cite{GelmanRubin} of $R-1 < 0.01$.
\begin{table}
\centering
\begin{tabular}{|l|cc|}
\hline
Parameter & Mean $B_\lambda=0$& Mean \\
\hline
$\omega_{\text b}$ & $0.0222^{+0.0011}_{-0.0010}$
& $0.0222 \pm 0.0010$ \\
$\omega_{\text c}$ & $0.109^{+0.010}_{-0.009}$ & $0.109 \pm 0.010$\\
$\theta$ & $1.040^{+0.004}_{-0.005}$ & $1.040^{+0.004}_{-0.005}$\\
$\tau_{\rm dec}$ & $0.086^{+0.030}_{-0.027}$ & $0.086 ^{+0.029}_{-0.030}$\\
$\log \left[ 10^{10} A_\text{S}\right]$ & $3.06^{+0.06}_{-0.07}$ & $3.05^{+0.07}_{-0.06}$ \\
$n_{\text S}$ & $0.956^{+0.024}_{-0.025}$ & $0.956^{+0.025}_{-0.026}$ \\
$B_{1 {\rm Mpc}}$/nG & ... & $<5.0 \, $  \\
$n_{\text B}$ & ... & $<-0.12 $  \\
\hline
$H_0/{\rm km}\,{\rm s^{-1}}{\rm Mpc}^{-1}$ & $71.5^{+4.6}_{-4.3}$ 
& $74.4^{+4.6}_{-5.4}$\\
\hline
\end{tabular}
\caption{\label{tab_data}
Mean parameter values and bounds of the central 95\%-credible intervals without (left column) 
and with (right column) PMF. Note that $H_0$ is a derived parameter.}
\end{table}

\begin{table}
\centering
\begin{tabular}{|l|c|c|c|}
\hline
Instrument & \multicolumn{1}{|c|}{LFI} &\multicolumn{2}{|c|}{HFI} \\
\hline
Center frequency GHz & 70 & 100 & 143  \\\hline
Mean FWHM (arcmin)  & 13.0 &  9.6 & 7.0 \\
\hline
$\Delta T/T$ per FWHM$^2$ (Stokes $I$) & 4.45 & 2.12 & 1.56 \\
\hline
$\Delta T/T$ per FWHM$^2$ (Stokes $Q\& U$) & 6.29 &  3.39 & 2.90 \\
\hline
\end{tabular}
\caption{\label{PlanckPer}
Planck channels considered in this analysis and relative performances in the extended mission (four surveys).}
\end{table}

\begin{table}
\centering
\begin{tabular}{|l|cc|}
\hline
Parameter & Input value & Mean \\
\hline
$\omega_{\text b}$ & $0.0227$ & $0.0227 \pm 0.0003$ \\
$\omega_{\text c}$ & $0.108$ & $0.108^{+0.003}_{-0.002}$ \\
$\theta$ & ... & $1.040 \pm 0.001 $ \\
$\tau_{\rm dec}$ & $0.089$ & $0.089 ^{+0.010}_{-0.008}$ \\
$\log \left[ 10^{10} A_\text{S}\right]$ & $3.1$ & $3.08^{+0.02}_{-0.01}$ \\
$n_{\text S}$ & $0.960$ & $0.961 \pm 0.008$ \\
$B_{1 {\rm Mpc}}$/nG &... & $< 2.7 $\\
$n_{\text B}$ & ... & $<-0.05$  \\
\hline
$H_0/{\rm km}\,{\rm s^{-1}}{\rm Mpc}^{-1}$ & $72.4$ & $72.3 \pm 1.3$ \\
\hline
\end{tabular}
\caption{\label{tab_planck}
Input and mean parameter values with bounds of the central 95\%-credible intervals for the Planck simulated data. 
Note that $H_0$ is a derived parameter.}
\end{table}

In Table \ref{tab_data} are reported the results of our analysis on current CMB data.
We compare the results for the six standard cosmological parameters both considering or not 
the PMF contribution. The comparison shows that neither the means 
nor the bounds of the cosmological parameters of the $\Lambda$CDM model with reionization are basically 
affected by the presence of PMF: this means that $B_\lambda$ 
and $n_B$ are not degenerate with the other six parameter of the 
concordance cosmological model. 

In Figs. \ref{Data_BnonB}  and  \ref{Data_tri} we show the plots of the MCMC results.
We derive the following constraints on the amplitude and spectral index of PMF: 
$B_{1 {\rm Mpc}}<5.0$ nG and $n<-0.12$ at $95\% $ confidence level.

In Fig. \ref{Data_tri} we compare the results obtained including or excluding the scalar contribution.
We note how, as expected from previous results on CMB anisotropies \cite{yamazaki,PFP,FPP}, the scalar contribution has a very 
little impact on the constraints on the amplitude $B_\lambda$ and spectral index $n_B$. 
This is due to the different shape of the CMB temperature spectra induced by the scalar 
and vector contribution of a stochastic background 
of PMF: compared to the CMB anisotropies sourced by the adiabatic mode, 
the PMF scalar mode is the dominant contribution on large scales, whereas 
the vector is the important one at high multipoles.
Since the PMF contribution does not suffer of the Silk damping, only 
the vector contribution has the main chance to affect the CMB 
temperature sourced by the standard adiabatic mode. 

A stochastic background of PMF has also an impact on the matter power spectrum as shown in \cite{FPP,SL}.
In order to investigate if the addition of matter power power 
spectrum data could improve the 
constraints on the PMF, we perfomed
a MCMC analysis adding the large scale structure (LSS) data of SDSS LRG DR4 \cite{sdss} to the 
CMB anisotropy data used before. 
We used the DR4 release because the DR7 requires
the use of non-linear tools to compute the matter power spectrum which are unavailable for $\Lambda$CDM plus PMF.
We have verified that the addition of the SDSS data does not introduce any significant improvement on the constraints
on PMF amplitude. The reason for the lack of improvement is that the matter power spectrum data reach scales of the order of 
$0.2\, h {\rm Mpc}^{-1}$ where only PMF with positive spectral index have an impact, 
but PMF with blue spectral indices are already strongly constrained with CMB anisotropy data,
therefore the results do not show any significant improvement. This result of a negligible impact of SDSS DR4 data with respect 
to current CMB data is in agreement with \cite{shawlewis2}. \\

Let us compare our constraints with other investigations.
After our manuscript appeared on the archive, 
Shaw and Lewis obtained similar CMB constraints - of the order of  $6$ nG at $95\%$ confidence level - 
with similar data sets in Ref. \cite{shawlewis2}. 
In \cite{yamazakilast} stronger constraints of the order of $3$ nG at $95\% $ confidence level were reported.
The reason for slightly different result might be either in the different datasets or in the different methodology used. 
Concerning the use of matter power spectrum, 
the present manuscript and Ref. \cite{shawlewis2} agree that LSS data add very little to CMB current constraints, 
when only linear scales are correctly included: 
we obtain analogous results either with SDSS DR4 \cite{sdss} or 2dF \cite{2dF}, without changing the CMB constraints in Table I.
Another 
difference might be in the use of CBI data \cite{CBI} in \cite{yamazakilast}, whose foreground contamination 
at high $\ell$ has been stressed in Ref. \cite{ACBAR}: we believe that the difference might be due to either 
methodology and/or to the different use of CMB data. 

Let us end this section by commenting on the constraints on the magnetic spectral index 
we derive, i.e. $n_B < -0.12$ at $95\% $ confidence level. This would imply that positive values for $n_B$ 
are disfavoured and so would be causal mechanism which would have produced such PMF \cite{DC}.
Certainly there is a trend in $B_{\lambda} \,, n_B$ which allows for larger $B_{\lambda}$ for negative values of 
$n_B$: if we fix $n_B$ to the causal value of $2$ we obtain a much stronger constraint on the amplitude, 
i.e. $B_{1 {\rm Mpc}} < 0.023$ nG $95\% $ confidence level. 
At the same time, we have verified that for undectatable values by CMB of $B_{1 {\rm Mpc}}$ - fixed to $10^{-6}$ nG -  
no constraints are derived for $n_B$, as expected. The above results are obtained by the assumption of a linear prior on 
$B_{1 {\rm Mpc}}$. We have also performed just for comparison 
cosmological parameter extraction by sampling ${\rm Log}_{10} [B_{1 {\rm Mpc}}/(10 {\rm nG})]$ in the range $[-7,1]$
\cite{comment} 
and we show the posteriors in Fig. \ref{PlanckLog}. 
Since the log prior overweights the importance of small amplitudes for $B_{1 {\rm Mpc}}$ compared to the linear prior, 
the constraints obtained on $B_{1 {\rm Mpc}}$ by sampling 
logarithmically are tigher - $1$ nG at $95\% $ confidence level - , but depend strongly on the interval. 
As byproduct of the importance given to small amplitude for $B_{1 {\rm Mpc}}$ in sampling logarithmically, 
$n_B$ is unconstrained. Similar prior issues occur for other important cosmological parameters 
which are under the threshold of detection, as the amplitude of 
gravitational waves in standard inflationary scenarios \cite{VKH}.  


\section{Forecasts for {\sc Planck}}

In the perspective of the forthcoming {\sc Planck} \cite{bluebook} data we performed a MCMC analysis using  
simulated {\sc Planck} data. 
We created the mock data considering the updated angular resolution and sensitivities \cite{mandolesi,lamarre} 
for the extended and approved mission duration of four sky surveys, which are summarized in Table \ref{PlanckPer}.
We used the combination of the three central frequency channels where the CMB is dominant with respect
to foreground contributions: $70$ GHz, $100$ GHz, $143$ GHz, which have been combined with the 
inverse noise variance weighting technique.
We used the $92$ \% of the sky and assumed only CMB neglecting all possible foreground contamination.
In Table \ref{tab_planck} we report the results for cosmological parameters with 
the input parameters of the fiducial model which we choose very close to the WMAP 7 yrs best fit model without PMF. 

In Fig. \ref{DataPlanck_tri} we show the comparison
between the parameters constrained by current CMB data and the ones which will be constrained by {\sc Planck}:
note the great improvement given by {\sc Planck} on the constraints on cosmological parameters in presence of PMF.
We forecast the following constraints by {\sc Planck}: $B_{1 {\rm Mpc}} <2.7$ nG and $n_B<-0.054$ at $95\%$ confidence level.
The improvement obtained by {\sc Planck} alone on the constraint on $B_\lambda$
is mainly due to the better measurement of the CMB temperature anisotropies at high multipoles. 
The factor of $2$ improvement in the constraints on $B_\lambda$ reached by {\sc Planck} corresponds to a factor $16$ 
improvement in the characterization of $C_\ell$.

\section{Conclusions}

We studied the constraints on a stochastic background of PMF by current and forthcoming CMB data. 
In doing this, we have improved the theoretical CMB predictions in presence of a stochastic background of PMF. 
We gave approximations for the relevant components of the EMT for any spectral index $n_B$. We considered 
the correlation between $\rho_B$ and $L_B$ and showed how the previous choice of total 
anti-correlation \cite{FPP,PFP} was a rather good approximation, in particular for red values of $n_B$.
We computed the initial conditions for magnetic scalar mode in presence of matter corrections 
showing how these corrections do not affect the results
contrary to what claimed in \cite{Kojima}.

On the basis of previous works, we have considered only the scalar and vector contribution, and by using their regular 
initial condition, 
we constrain $B_{1 {\rm Mpc}} < 5.0$ nG and $n_B < - 0.12$ at $95\%$ confidence level.
with the most updated combination of CMB anisotropies. 
{\sc Planck} will be able to constrain the spectrum of a stochastic background of PMF even further at the level of $2.7$ nG.

\vspace{1cm}

\begin{center}
{\bf Acknowledgements}
\end{center}
We wish to thank C. Caprini, R. Durrer, J. Hamann and L. Hollenstein for useful comments. 
This work has been done in the framework of the {\sc Planck} LFI activities and 
is partially supported by ASI contract {\sc Planck} LFI 
activity of Phase E2. We acknowledge the use of the Legacy Archive for Microwave 
Background Data Analysis (LAMBDA). Support for LAMBDA 
is provided by the NASA Office of Space Science.
\begin{widetext}

\begin{figure}
\includegraphics[scale=1]{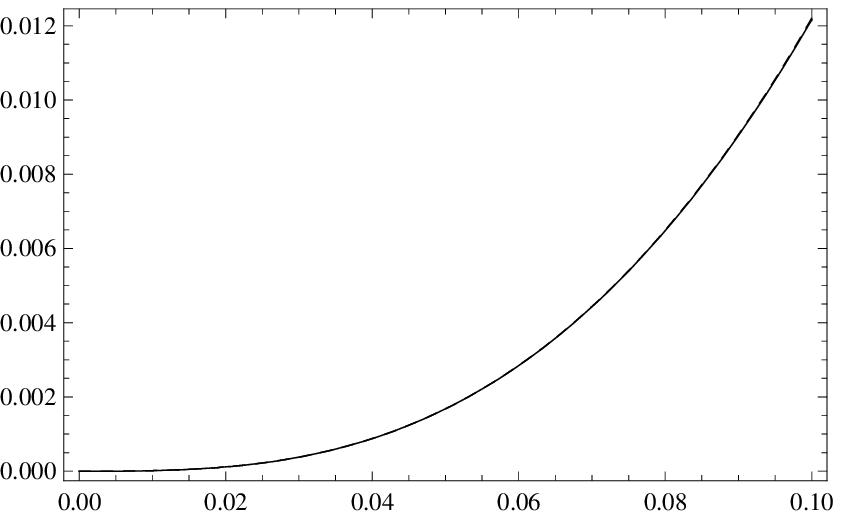}\includegraphics[scale=1]{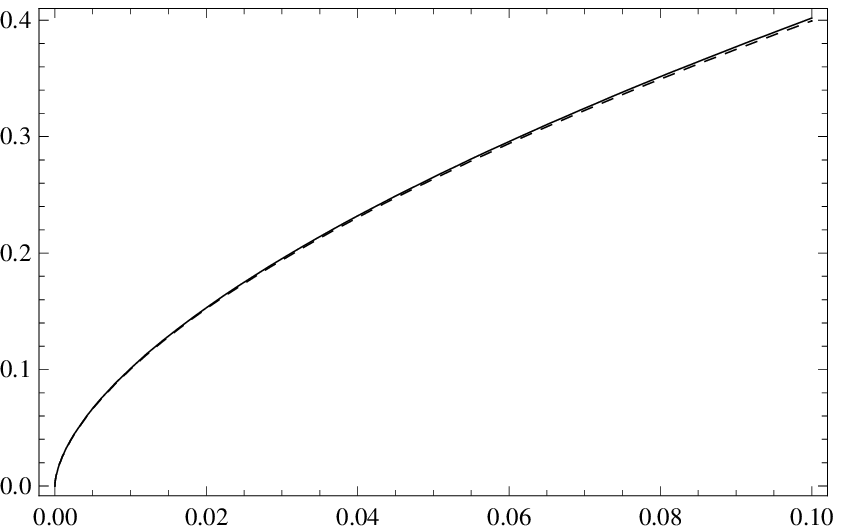}\\
\includegraphics[scale=1]{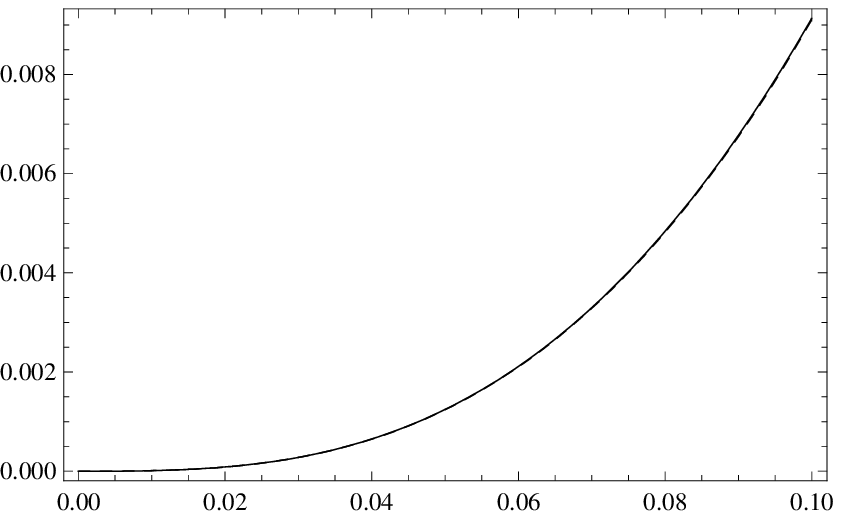}\includegraphics[scale=1]{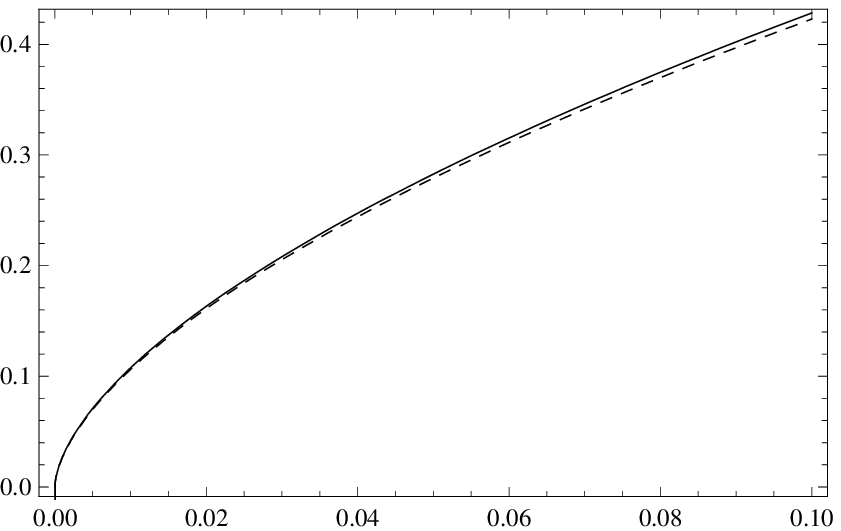}\\
\includegraphics[scale=1]{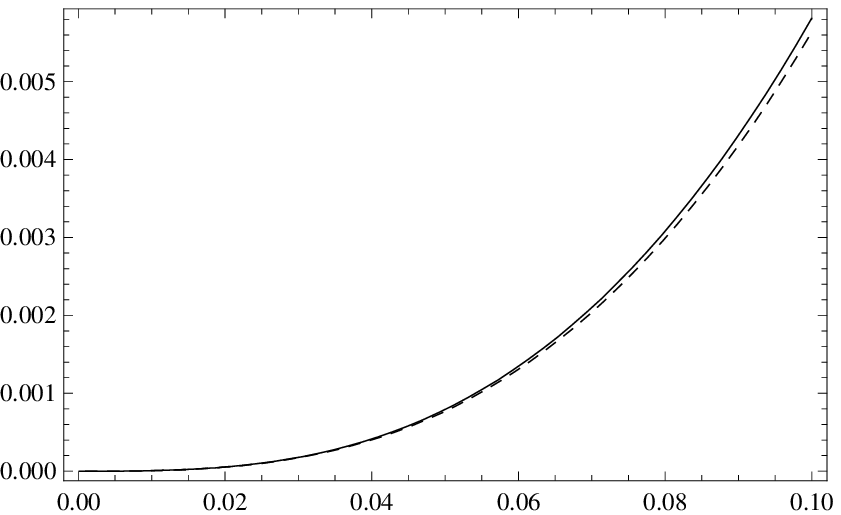}\includegraphics[scale=1]{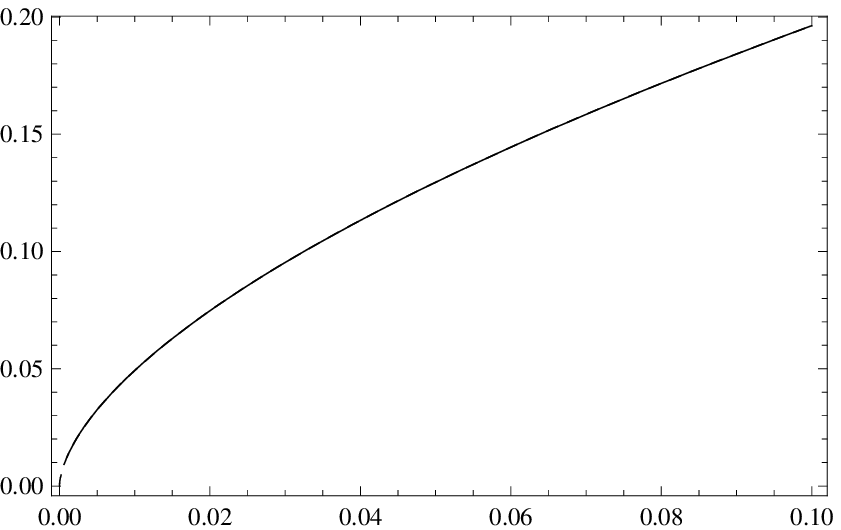}
\caption{Comparison of spectral fit (dashed line) and exact spectrum (solid line) for the rescaled magnetic energy density $|\rho_B (k)|^2/U$ (top panel), scalar Lorentz force $|L_B (k)|^2/U$ (middle panel) and the vector part of the anisotropic stress $|\Pi^{(V)} (k)|^2/(2 U)$ (bottom panel) as a function of $k/k_D$ with $U=(A^2 k_D^{2n_B+3})/(512 \pi^4) $. 
In the left (right) column $n_B=2.3$ ($n_B=-2.7$) is displayed.}
\label{Comparison_fit_scal}
\end{figure}

\appendix

\section{PMF EMT spectral fits}
Since the spectral behaviour is rather complicated, in order to have the best fit possible we decided 
to divide the fits for spectral index ranges.
Together with the natural separation between indices smaller and greater than $-3/2$, 
which is necessary due to the completely different behaviour, we added a 
further separation between negative and positive spectral indices.
The leading terms remain the same as in the infrared limit of the exact spectra: 
white noise for blue indices and infrared dominated
for the red ones.

\subsection{Scalar Spectra}
{\bf $n_B \ge 0$}
\be
|\rho_B(k,n_B)|^2_{fit}= \frac{A^2 k_D^{2n_B+3}}{512 \pi^4 k_*^{2n_B}} \Big(\frac{4}{2n_B+3}-\tilde{k}+\sum_{i=1}^3 A_i \tilde{k}^{i+1} + A_4 
\tilde{k}^{(2 n_B + 3)}\Big)
\ee
{\bf $-3/2<n_B<0$}
\be
|\rho_B(k,n_B)|^2_{fit} = \frac{A^2 k_D^{2n_B+3}}{512 \pi^4 k_*^{2n_B}}\Big(\frac{4}{2n_B+3}-\tilde{k}+\sum_{i=1}^3 B_i \tilde{k}^{i+1}
+B_4 \tilde{k}^{(2 n_B + 3)}\Big)
\ee
{\bf $-2.9<n_B<-3/2$}
\be
|\rho_B(k,n_B)|^2_{fit} = \frac{A^2 k_D^{2n+3}}{512 \pi^4 k_*^{2n_B}}
\Big(\frac{4}{2n_B+3}-\tilde{k}+C_1 \tilde{k}^{(2 n_B + 3)}\Big)
\ee
{\bf Coefficients}\newline
For positive spectral indexes:
\ba
A_1&=&-0.8998 -\frac{0.03926}{n_B}+1.419 n_B-0.695 n_B^2+0.2642 n_B^3-0.05418 n_B^4+0.004595 n_B^5\nonumber\\
A_2&=&0.3265+\frac{0.0008383}{n_B} +0.01671 n_B - 0.1016 n_B^2 + 0.00989 n_B^3 -0.002607 n_B^4 + 0.0002657 n_B^5\nonumber\\
A_3&=&11.3 - \frac{1.631}{n_B} - 21.8 n_B + 19.66 n_B^2 - 9.243 n_B^3 + 2.184 n_B^4 -0.2041 n_B^5\nonumber\\
A_4&=&0.3919 + \frac{0.3111}{n_B}-5.899 n_B + 9.607 n_B^2 -6.21 n_B^3 + 1.79 n_B^4 - 0.1918 n_B^5\nonumber
\ea
for the negative spectral indexes we have:
\ba
B_1&=&\frac{1}{5} (-825 - 2848 n_B - 3980 n_B^2 - 2490 n_B^3 - 580 n_B^4) - \frac{57}{5 n_B}\nonumber\\
B_2&=&\frac{1}{50} (15 - 4 n_B^2)\nonumber\\
B_3&=&\frac{1}{25} (-5 - 11 n_B - 8 n_B^2 - 3 n_B^3)\nonumber\\
B_4&=&\frac{171}{25 n_B} + \frac{1}{50} (4673 + 12900 n_B + 11500 n_B^2 + 1950 n_B^3 - 1155 n_B^4)\nonumber
\ea
for strongly negative:
\be
C_1=-\frac{10527877}{200 n_B}+\frac{-126773640 - 114087370 n_B - 39615180 n_B^2 +4157430 n_B^3 + 7369110 n_B^4 + 2081486 n_B^5 + 198571 n_B^6}{1000}
\ee

\subsection{Lorentz Force Spectra}
{\bf Positive $n_B$}
\be
|L(k,n_B)|^2_{fit} = \frac{A^2 k_D^{2n_B+3}}{512 \pi^4 k_*^{2n_B}}
\Big(A^L_1-\frac{2}{3}\tilde{k}+A^L_2 \tilde{k}^2+A^L_3 \tilde{k}^{2 n_B + 3}\Big)
\ee
{\bf $-3/2<n_B<0$}
\be
|L(k,n_B)|^2_{fit} = \frac{A^2 k_D^{2n_B+3}}{512 \pi^4 k_*^{2n_B}}
\Big(B^L_1-\frac{2}{3}\tilde{k}+B^L_2 \tilde{k}^{2 n_B + 3}\Big)
\ee
{\bf $-2.9<n_B<-3/2$}
\be
|L(k,n_B)|^2_{fit} = \frac{A^2 k_D^{2n_B+3}}{512 \pi^4 k_*^{2n_B}}
\Big(C^L_1-\frac{2}{3}\tilde{k}+C^L_2 \tilde{k}^{2 n_B + 3}\Big)
\ee

{\bf Coefficients}\newline
For positive spectral indexes:
\ba
A^L_1&=&0.933635 +\frac{0.00460612}{n_B}-0.505278 n_B+0.183487 n_B^2-0.0238037 n_B^3 - 0.00985191 n_B^4 +0.00437658 n_B^5 - 
0.000504247 n_B^6\nonumber\\
A^L_2&=&0.22309-\frac{0.021189}{n_B}-0.152155 n_B+0.427087 n_B^2-0.184484 n_B^3-0.0111374 n_B^4+0.0292611 n_B^5-0.00571069 n_B^6\nonumber\\
A^L_3&=&1.84015 - \frac{0.319013}{n_B}-3.60452 n_B+2.88574 n_B^2-0.797507 n_B^3-0.145007 n_B^4+0.116527 n_B^5-0.0163659 n_B^6\nonumber
\ea
for the negative spectral indexes we have:
\ba
B^L_1&=&\frac{1}{100} (1630 + 4240 n_B + 3360 n_B^2 - 2080 n_B^3 - 1960 n_B^4 + 1970 n_B^5 +1559 n_B^6) + \frac{41}{25 n_B}\nonumber\\
B^L_2&=&\frac{1}{100} (-854 - 2838 n_B - 2710 n_B^2 + 1390 n_B^3 + 1705 n_B^4 - 1530 n_B^5 -1340 n_B^6) -\frac{4}{5 n_B}\nonumber
\ea
for strongly negative:
\ba
C^L_1&=&\frac{1}{50} (1327860 + 1077425 n_B + 321980 n_B^2 - 50935 n_B^3 - 60380 n_B^4 -15115 n_B^5 - 1302 n_B^6) + \frac{60569}{5 n_B}\nonumber\\
C^L_2&=&-\frac{241194}{5 n_B} 
+ \frac{(-117123100 - 106256700 n_B - 37275000 n_B^2 + 3787200 n_B^3 + 6930290 n_B^4 + 1971640 n_B^5 + 189111 n_B^6)}{1000}\nonumber
\ea

\subsection{Vector Spectra}
{\bf Positive $n_B$}
\be
|\Pi^{(V)}(k,n_B)|^2_{fit} = \frac{A^2 k_D^{2 n_B +3}}{256 \pi^4 k_*^{2n_B}}
\Big(A^V_1-\frac{5}{12}\tilde{k}+A^V_2+A^V_3 \tilde{k}^{2 n_B + 3}\Big)
\ee
{\bf $-3/2<n_B<0$}
\be
|\Pi^{(V)}(k,n_B)|^2_{fit} = \frac{A^2 k_D^{2 n_B +3}}{256 \pi^4 k_*^{2n_B}}
\Big(B^V_1-\frac{5}{12}\tilde{k}+B^V_2 \tilde{k}^2+B^V_3 \tilde{k}^{2 n_B + 3}\Big)
\ee
{\bf $-2.9<n_B<-3/2$}
\be
|\Pi^{(V)}(k,n_B)|^2_{fit} = \frac{A^2 k_D^{2 n_B +3}}{256 \pi^4 k_*^{2n_B}}
\Big(C^V_1-\frac{5}{12}\tilde{k}+C^V_2 \tilde{k}^{2 n_B + 3}\Big)
\ee
{\bf Coefficients}\newline
For positive spectral indexes:
\ba
A^V_1&=&\frac{29500 - 16100 n_B + 5850 n_B^2 - 765 n_B^3 - 314 n_B^4 + 140 n_B^5 -16 n_B^6}{50000}\nonumber\\
A^V_2&=&\frac{-845 + 2600 n_B - 690 n_B^2 + 124 n_B^3}{10000}\nonumber\\
A^V_3&=&\frac{1}{500} (-280 + 545 n_B - 425 n_B^2 + 112 n_B^3)\nonumber
\ea
for the negative spectral indexes we have:
\ba
B^V_1&=&\frac{26}{25 n_B}+\frac{1}{100} (1040 + 2698 n_B + 2140 n_B^2 - 1327 n_B^3 - 1249 n_B^4 + 1255 n_B^5 +992 n_B^6)\nonumber\\
B^V_2&=&\frac{1}{100} (-2192 - 4681 n_B - 2132 n_B^2 + 2235 n_B^3 + 908 n_B^4 - 1464 n_B^5 -744 n_B^6) -\frac{53}{20 n_B}\nonumber\\
B^V_3&=&\frac{73}{50 n_B} +\frac{1}{100} (1078 + 1616 n_B - 243 n_B^2 - 735 n_B^3 + 471 n_B^4 + 59 n_B^5 -342 n_B^6)\nonumber
\ea
for strongly negative:
\ba
C^V_1&=&\frac{445985}{500 n_B} + \frac{(19923100 + 16525360 n_B + 5113265 n_B^2 - 742742 n_B^3 - 956890 n_B^4 -246837 n_B^5 - 
21843 n_B^6}{1000}\nonumber\\
C^V_2&=&\frac{-29003653 - 25196700 n_B - 8371900 n_B^2 + 995460 n_B^3 + 1561850 n_B^4 + 429404 n_B^5 + 40254 n_B^6}{1000} - 
\frac{124807}{10 n_B}\nonumber
\ea
\section{Cross Correlators Exact Solutions}
\begin{figure}
\begin {tabular} cc
\includegraphics[scale=1]{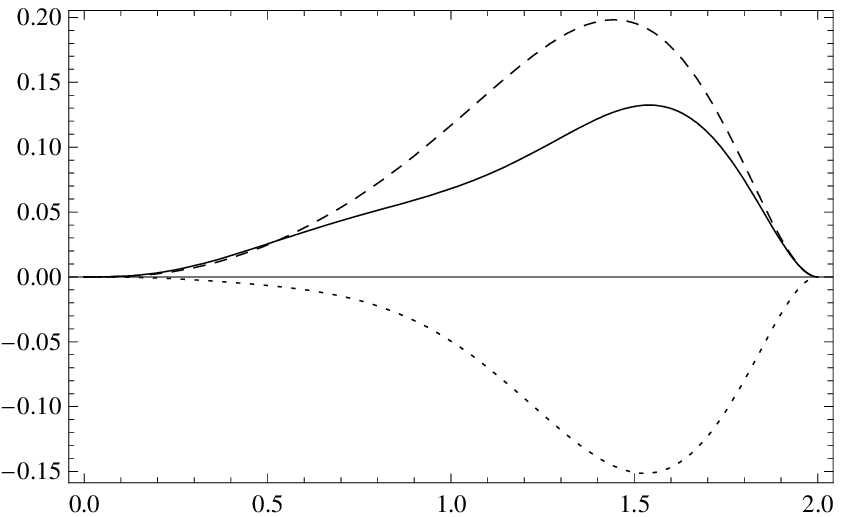}
\includegraphics[scale=1]{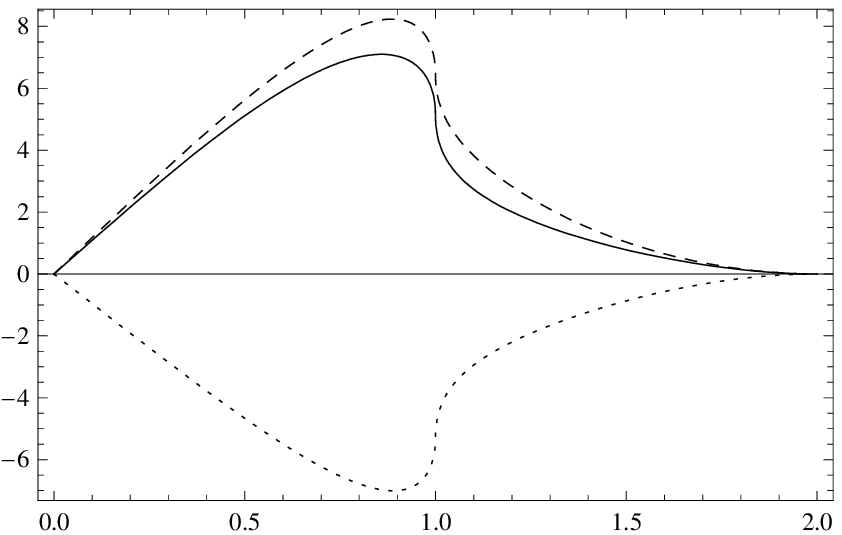}
\end{tabular}
\caption{The cross-correlator $k^3 \langle\rho_B(k)L_B(k)\rangle/U$ (dotted line) for $n_B=2$ and $n_B=-5/2$ is plotted versus $k/k_D$ 
in comparison with $k^3 |\rho_B (k)|^2/U$ (solid line) and $k^3 |L_B (k)|^2/U$ (dashed line) 
with $U=(A^2 k_D^{2n_B+3})/(512 \pi^4) $.}
\label{Correlation_k}
\end{figure}
\begin{figure}
\includegraphics[scale=0.5]{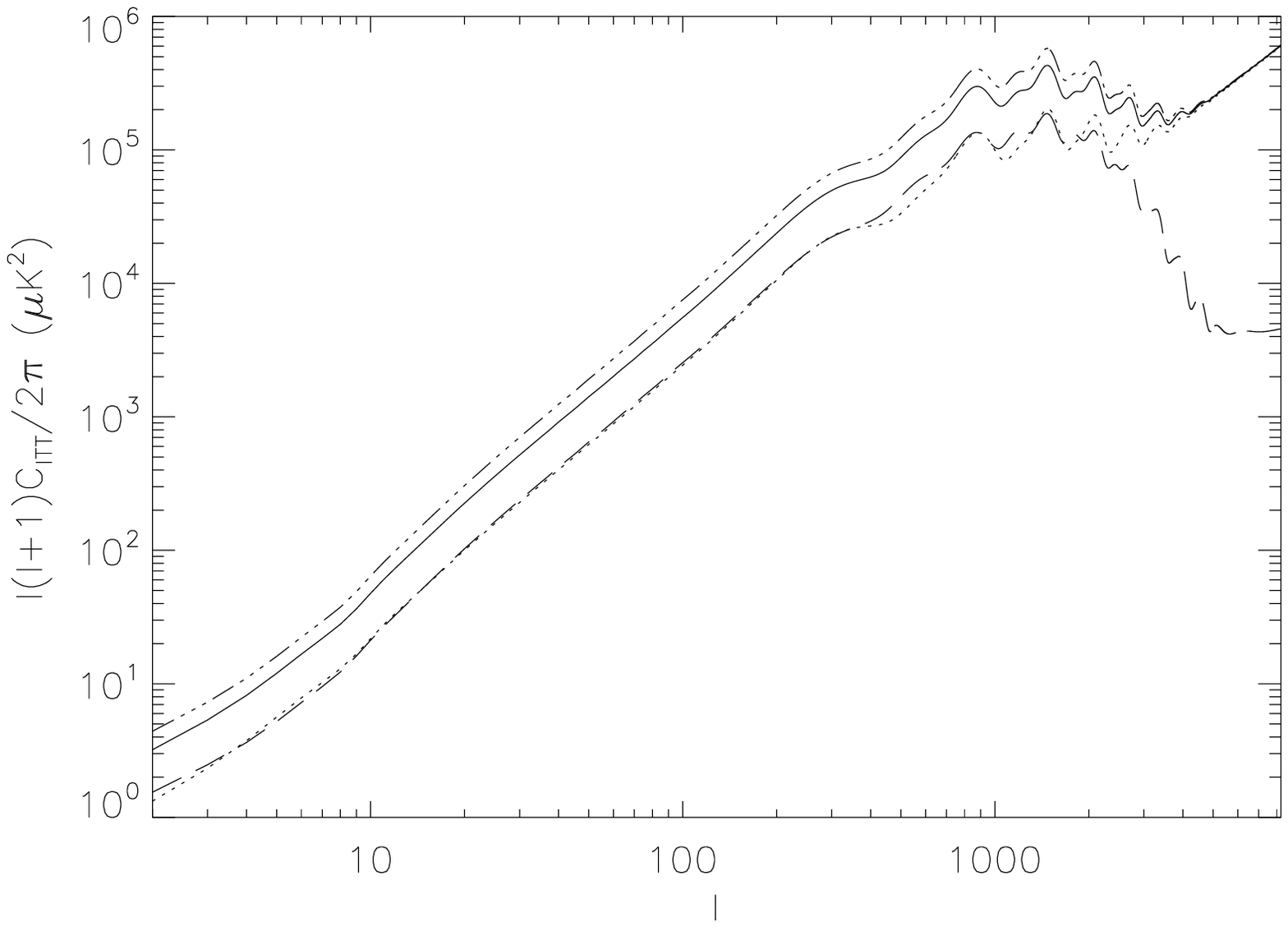}
\includegraphics[scale=0.5]{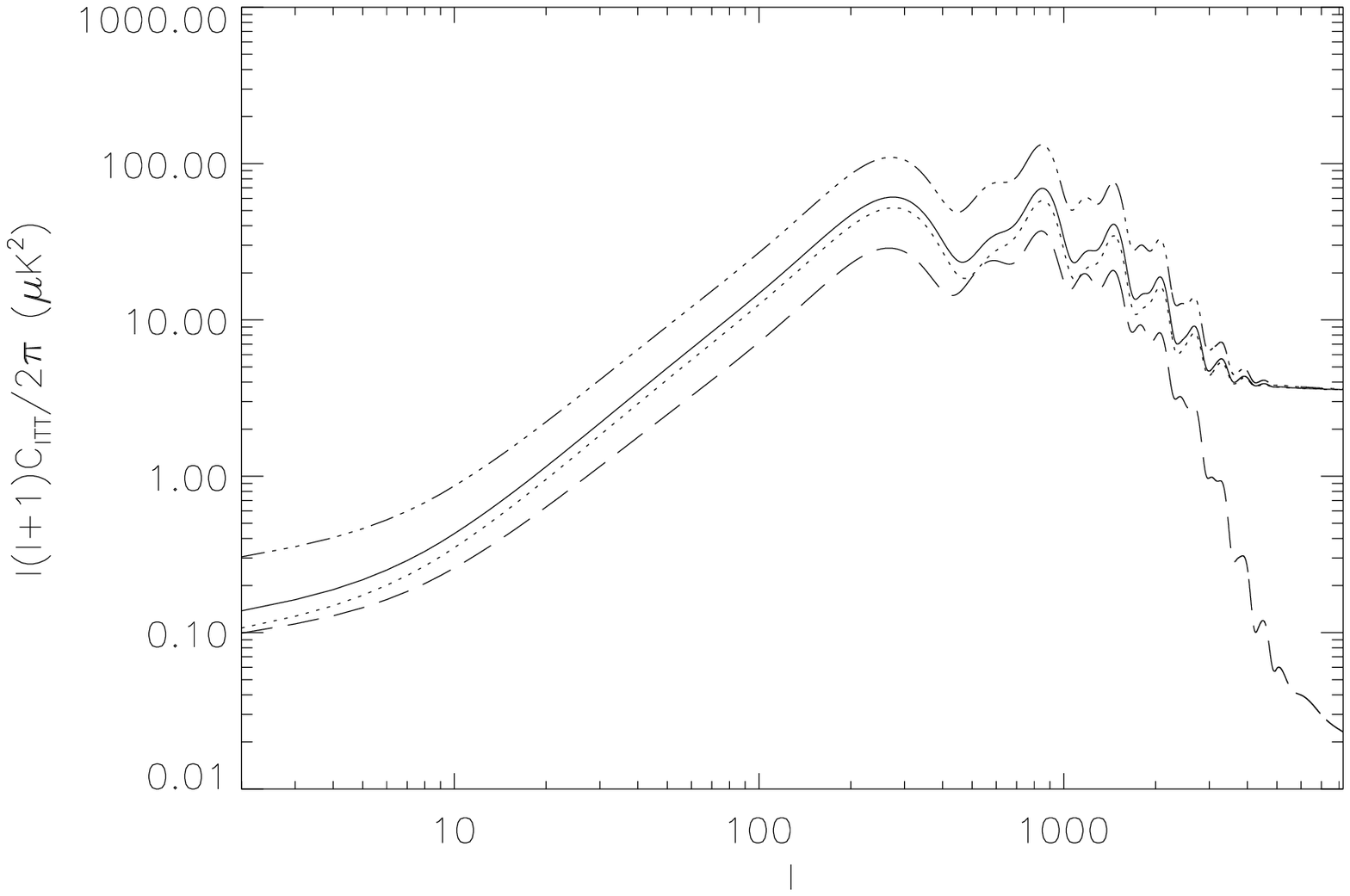}
\caption{We computed the scalar power spectrum with the contribution of the Lorentz force-energy density cross correlation 
for $n_B=2$ (top panel) and $n_B=-2.5$ (bottom panel). 
Solid line represents the correct assumption on the cross-correlation, the dotted line represents the result considering the 
Lorentz force and the energy density fully anti-correlated, triple dotted-dashed line represents the uncorrelated sum 
and the dashed line represents the result assuming full correlation. 
}
\label{Correlation_CMB}
\end{figure}
To give the reader an idea of the complete analytical form of the Lorentz force and energy density cross correlator
we show the exact results for two 
representative spectral indices $n_B=2$ and $n_B=-2.5$:
\ba
\langle\rho_B(k)L_B(k)\rangle|_{n_B=2}&=&\frac{A^2 k_D^7}{1024\pi^5 a^8}\Big[-\frac{4}{21}+\frac{\tilde{k}}{2} 
-\frac{8 \tilde{k}^2}{15} +\frac{\tilde{k}^3}{6} +\frac{\tilde{k}^5}{96} - \frac{3\tilde{k}^7}{1120}\Big]\nonumber\\
\langle\rho_B(k)L_B(k)\rangle|_{n_B=-5/2}&=&\frac{A^2}{1024\pi^5 a^8 k_D^2}\Big[\frac{16(4 - 65\tilde k + 
59\tilde k^2 - 2\tilde k^3 + 4\tilde k^4)}{(105 \sqrt{|1 -\tilde k|}\tilde k^3)} -\frac{64 + 448 \tilde k^2 + 42 \tilde k^4}{105 \tilde k^3}\Big]
\ea
where $\tilde k=k/k_D$. In Fig. \ref{Correlation_k} we show
the comparison between the energy density the Lorentz force
and their cross correlation for $n_B=-2.5$ and $n_B=2$: the cross-correlation is negative in the whole range of scales.
In Fig. \ref{Correlation_CMB} we show the impact of various assumptions for the 
cross-correlation between the magnetic energy density and 
the Lorentz force on the power spectrum of CMB temperature anisotropies.
\section{Scalar Initial Conditions with Matter Corrections}
We computed the initial conditions for magnetized scalar perturbations for the leading 
regular growing mode which is the one which contributes to CMB anisotropies, 
we do not consider any isocurvature or decaying modes.
Initial conditions are the solution of the coupled Einstein 
Boltzmann equation system with PMF contributions on long wavelenghts at early epochs.
The Einstein equations with the contribution of PMF in the synchronous
gauge (with the notation of  \cite{MB}) are:
\begin{eqnarray}
&&k^2 \eta-\frac{1}{2} {\mathcal{H}} \dot h = - 4\pi G a^2 (\Sigma_n\,\rho_n\delta_n+\rho_B)  \,,\nonumber \\
&&k^2 \dot\eta = 4\pi G a^2 \Sigma_n(\rho_n+P_n)\theta_n \,, \nonumber\\  
&&\ddot h +2 {\mathcal{H}} \dot h -2 k^2 \eta = - 8\pi G a^2\Big(\Sigma_n \,c_{s \, n}^2\rho_n\delta_n+\frac{\delta\rho_B}{3}\Big) \nonumber \,, \\
&&\ddot h+6\ddot \eta +2 {\mathcal{H}} (\dot h+6 \dot\eta)-2 k^2\eta
= -24 \pi G a^2  [\Sigma_n(\rho_n+P_n) \sigma_n+\sigma_B] ,
\label{Einsteineqs}
\end{eqnarray}
where $n$ represents the various species of the plasma: $b$ for baryons, $c$ for cold dark
matter (CDM), $\gamma$ for photons and $\nu$ for massless neutrinos.
The scalar metric perturbations are represented by the two scalar potential $h,\,\eta$
while fluid perturbations are represented by $\delta_i,\,\theta_i,\,\sigma_i$ which are respectively the
density contrast the fluid velocity and the anisotropic stress in the notation of \cite{MB}.
The overdots denote the derivative with respect to conformal time.
On long wavelenght we can expand the metric and fluid perturbations 
in series of $k\tau<<1$.
In our previous works \cite{FPP,PFP} we computed the intial conditions deep in the radiation era
 with the approximation of a universe dominated only by relativistic degrees of freedom (radiation and massless 
neutrinos). Usually considering $a(\tau) \propto \tau$ is a rather good approximation for setting the 
initial conditions (see however Ref. \cite{Kojima} for a different claim stressing the importance 
of matter corrections). 
We have therefore extended our previous results \cite{PFP} to the case in which the matter contribution 
is taken into account.
Considering the matter contribution in the initial condition needs the introduction of matter correction also in 
the evolution of the scale factor \cite{LewisCAMBnotes}:
\ba
a(\tau)&=&\frac{\Omega_m H_0^2}{\omega^2}\Big(\omega\tau+\frac{1}{4}\omega^2 \tau^2\Big)\,,\nonumber\\
\omega &=&\frac{\Omega_m H_0}{\sqrt{\Omega_\nu+\Omega_\gamma}}\,,\nonumber
\ea
where $\Omega_m=\Omega_c+\Omega_b$ and 
the Hubble parameter can be expanded as $\mathcal{H}\propto \frac{1}{\tau}+\frac{\omega}{4}$.
The Einstein equations become:
 \ba
&&k^2 \eta -\frac{1}{2}{\mathcal{H}}\dot h=-4\pi G a^2 (\rho_\gamma \delta_\gamma+ \rho_\nu\delta_\nu+\rho_b\delta_b+\rho_c\delta_c+(\rho_\nu+\rho_\gamma)\Omega_B) \,,\\
&&k^2 \dot \eta = 4\pi G a^2\left(\frac{4}{3}\rho_\gamma \theta_\gamma+\frac{4}{3}\rho_\nu \theta_\nu+\rho_b \theta_b\right)\,,\\
&&\ddot h +2 {\mathcal{H}} \dot h-2 k^2 \eta = - 8\pi G a^2 \left[\frac{1}{3}\rho_\gamma \delta_\gamma+\frac{1}{3}\rho_\nu  \delta_\nu+\frac{1}{3}(\rho_\gamma+\rho_\nu)  \Omega_B \right] \,,\\
&&\ddot h+6 \ddot \eta+2 {\mathcal{H}}(\dot h+6\dot\eta)-2k^2 \eta= -24 \pi G a^2\left[\frac{4}{3} \rho_\nu \delta_\nu+(\rho_\nu+\rho_\gamma)\Big(\frac{\Omega_B}{3}+L_B\Big)\right]\,,
\ea  
where PMF contributions are given in terms of their ratio with the fluid radiation density:
$\Omega_B=\rho_B/\rho_{rad}$, $L_B=L_B/\rho_{rad}$ and $\sigma_B=\sigma_B/\rho_{rad}$ (with $\rho_{rad}=\rho_\nu+\rho_\gamma$).
The initial conditions at leading order with PMFs which include the matter corrections are:
\begin{eqnarray}
h(k,\tau)&=&-\frac {3}{4} \Omega_B \omega \tau + \frac{9}{32} \Omega_B \omega^2 \tau^2 \nonumber\\
\eta(k,\tau)&=& \frac{1}{8} \Omega_B \omega \tau 
-\frac{3 \Omega_B \omega^2 \tau^2}{64}
+\frac{(-165L_B-55\Omega_B+28R_\nu \Omega_B)}{168(15+4R_\nu)}k^2 \tau^2\nonumber\\
\delta_\gamma(k,\tau)&=&-\Omega_B +\frac{\Omega_B \omega \tau}{2}
-\frac{3 \Omega_B \omega^2 \tau^2}{16}
-\frac{(3L_B+\Omega_B-R_\nu\Omega_B)}{6(-1 + R_\nu)}k^2 \tau^2 \nonumber\\
\delta_\nu(k,\tau)&=&-\Omega_B +\frac{\Omega_B \omega \tau}{2}-\frac{3 \Omega_B \omega^2 \tau^2}{16}
-\frac{(3L_B+\Omega_B-R_\nu\Omega_B)}{6R_\nu}k^2 \tau^2  \nonumber\\
\delta_b(k,\tau)&=&-3 \frac{\Omega_B}{4}+\frac{3 \Omega_B \omega \tau}{8} 
+\frac{1}{8} k^2 \tau^2 \Omega_B-\frac{9}{64} \Omega_B \omega^2 \tau^2 \omega^2 
-\frac{3 L_B k^2 \tau^2}{8(-1 + R_\nu)}\nonumber\\
\delta_c(k,\tau)&=&-\frac{3 \Omega_B}{4}+
\frac{3 \Omega_B \omega \tau}{8}-\frac {9}{64} \Omega_B \omega^2 \tau^2 \nonumber\\
\theta_\gamma(k,\tau)&=&\frac{3 L_B k^2 \tau}{4 (-1 + R_\nu)}
-\frac{\Omega_B}{4} k^2 \tau +k^2 \tau^2 
\left[-\frac{9 L_B(-1 + R_c)\omega}{16 (-1 + R_\nu)^2}
+\frac{(-4 + R_\nu + 3 R_c)\omega\Omega_B}{16(-1 + R_\nu)}\right]\nonumber\\
\theta_\nu(k,\tau)&=&\frac{3 L_B k^2 \tau}{4 R_\nu})
-\frac{k^2(-1+R_\nu) \Omega_B \tau}{4 R_\nu}+\frac{1}{16} k^2 \tau^2 \omega \Omega_B\nonumber\\
\theta_b(k,\tau)&=& \frac{3 L_B k^2 \tau}{4 (-1 + R_\nu)}-
\frac{1}{4} \Omega_B k^2 \tau +k^2 \tau^2 
\left[-\frac{9 L_B(-1 + R_c)\omega}{16 (-1 + R_\nu)^2}+\frac{(-4 + R_\nu + 3 R_c)\omega\Omega}{16(-1 + R_\nu)}\right]
\nonumber\\
\theta_c(k,\tau)&=& 0 \nonumber\\
\sigma_\nu(k,\tau)&=&-\frac{3 L_B + \Omega_B}{4 R_\nu} 
+\frac{\Omega_B k^2 (55 - 28R_\nu)\tau^2}{56 R_\nu (15 + 4 R_\nu)}
+\frac{165 L_B k^2 \tau^2}{56 R_\nu (15+4R_\nu)}\nonumber\\
F_3(k,\tau)&=&-\frac{3 k \tau (3 L_B + \Omega_B)}{14 R_\nu}
+\frac{165 L_B+55\Omega_B-28 R_\nu \Omega_B}{7(430 R_\nu+112R_\nu^2)}\,,
\end{eqnarray}
where $R_\nu=\rho_\nu/(\rho_\nu+\rho_\gamma)$.
We note that our results are in agreement, within our notation, with the one presented in \cite{SL}.
We verified that the inclusion of the matter correction in the initial conditions 
in our modified \texttt{CAMB} code does not produce any appreaciable
change in the results: $\Delta C_l/C_l\sim 10^{-5}$ which is the level of numerical noise, 
in contrast with Ref. \cite{Kojima}. 

\end{widetext}

\begin{widetext}{
\clearpage
\begin{figure}
\includegraphics[scale=0.55]{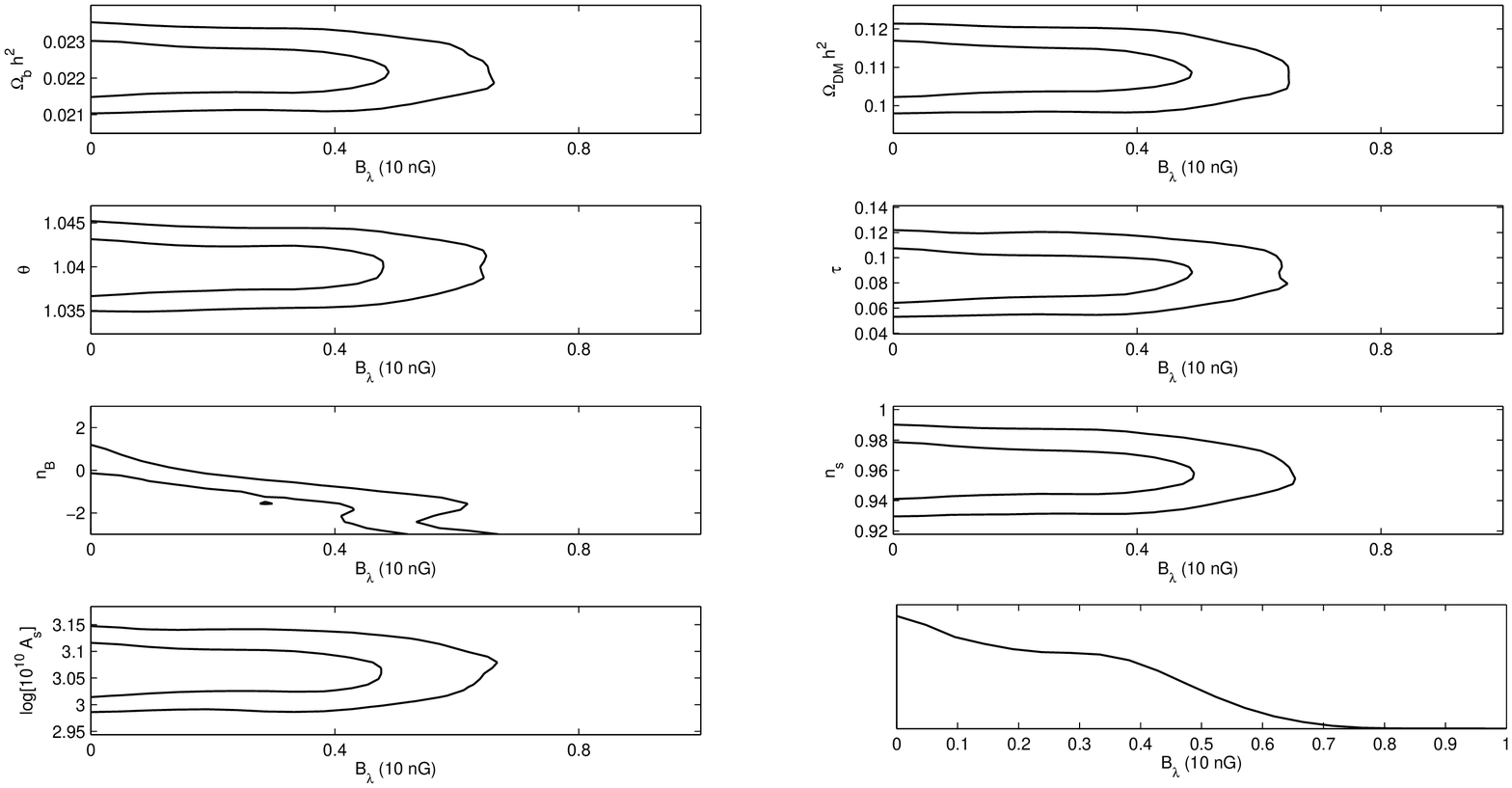}
\caption{Results of the MCMC constrained with WMAP 7 years, BICEP, QUaD and ACBAR
data. 
Curves are the $68\%$ and $95\%$ confidence level.  
Note that $B_\lambda$ (with $\lambda=1$ Mpc) is in $10$ nG units.}
\label{Data_BnonB}
\end{figure}
\begin{figure}
\includegraphics[scale=0.55]{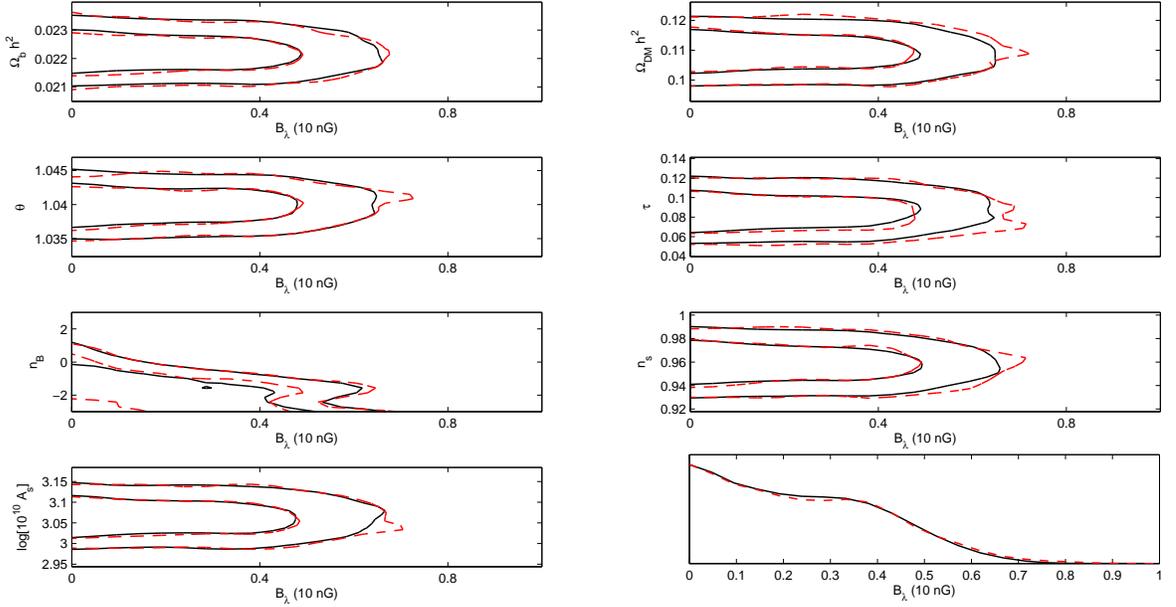}
\caption{Results of the MCMC constrained with WMAP 7 years, BICEP, QUaD and ACBAR
data.  
Curves are the $68\%$ and $95\%$ confidence level.
Solid lines are the results considering both scalar and vector magnetic modes while
dashed lines are the results taking into account
only the vector contribution. Note that $B_\lambda$ (with $\lambda=1$ Mpc) is in
$10$ nG units.}
\label{Data_tri}
\end{figure}
\clearpage
\begin{figure}
\includegraphics[scale=0.45]{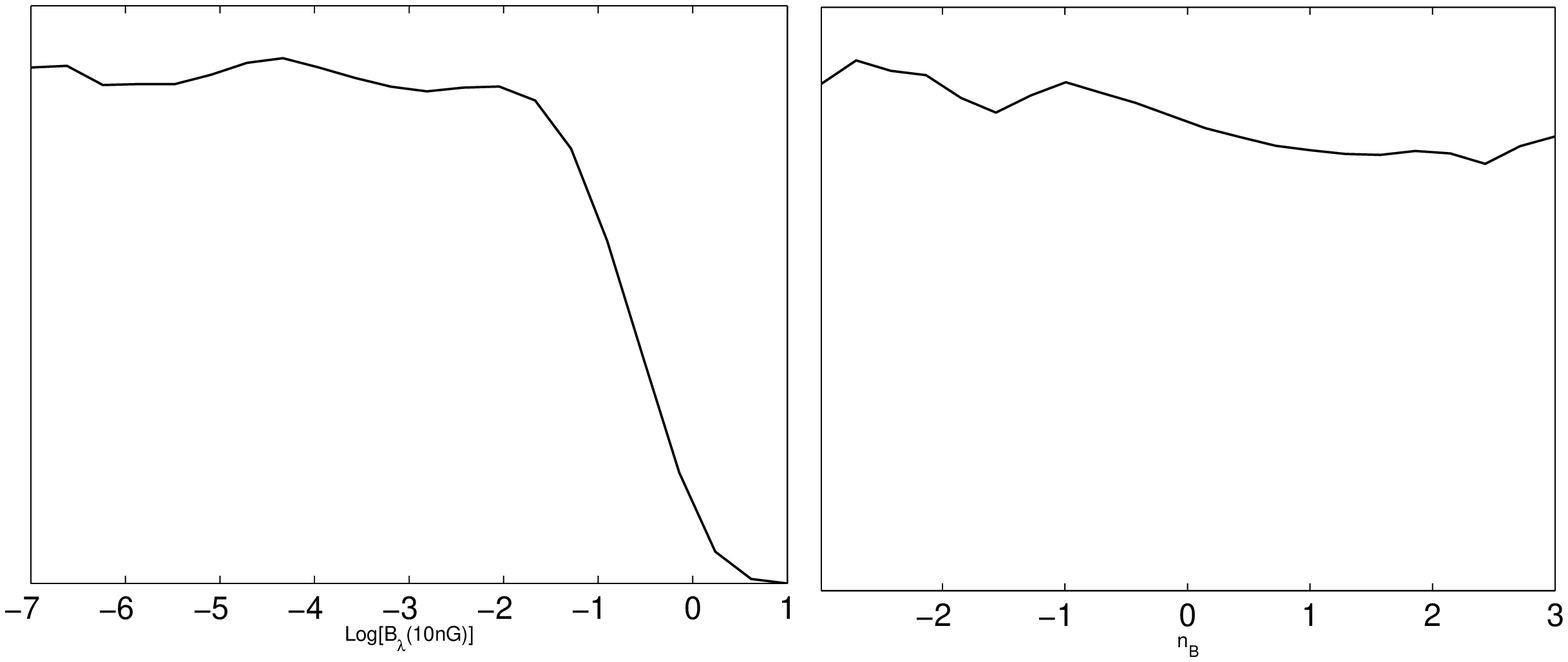}
\caption{Results of the MCMC for the PMF spectral index constrained with WMAP 7
years, BICEP, QUaD and ACBAR data and logarithmic prior for the PMF amplitude.
Curves are the $95\%$
and $68\%$ confidence level.}
\label{PlanckLog}
\end{figure}

\begin{figure}
\includegraphics[scale=0.55]{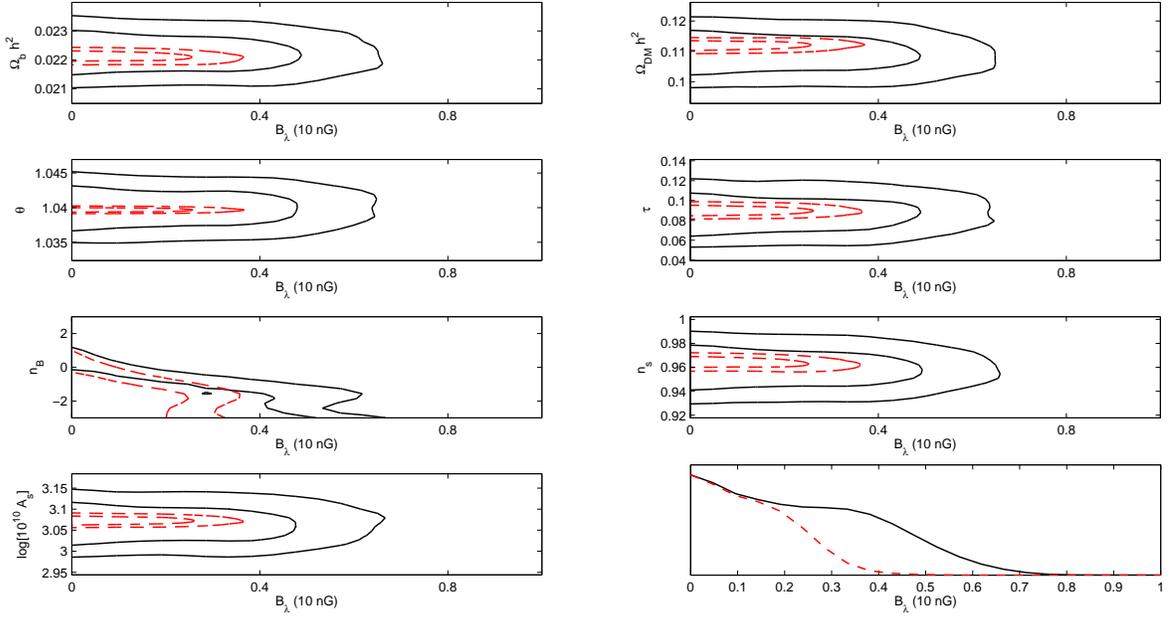}
\caption{Comparison of the results of the MCMC with real data used in Figs. 5, 6
(solid line)
and simulated {\sc Planck} data (dashed line). Curves are the $95\%$
and $68\%$ confidence level. Note that $B_\lambda$ (with $\lambda=1$ Mpc) is in $10$
nG units. Note that $k_D$ is a derived parameter.}
\label{DataPlanck_tri}
\end{figure}}

\end{widetext}
\clearpage

\end{document}